%% file: mainfile.tex
\documentclass[sigconf]{acmart}

\usepackage{amsthm}
\usepackage{amsmath}
 
\usepackage{amssymb}
\usepackage{float}
\usepackage{graphics}
\usepackage{graphicx}
\usepackage[skip=0pt]{caption}
\usepackage[skip=0pt]{subcaption}
\usepackage{algorithm}
\usepackage{algpseudocode}
\usepackage{bm}
\usepackage{color}
\usepackage{multirow}
\usepackage{makecell}
\usepackage{balance}
\usepackage{hyperref}

\newcommand{\myparatight}[1]{\smallskip\noindent{\bf {#1}:}~}
 
\allowdisplaybreaks

\graphicspath{ {figs/} }

\newcommand{\alg}{InferGuard}

\copyrightyear{2024}
\acmYear{2024}
\setcopyright{rightsretained}
\acmConference[WWW '24 Companion]{Companion Proceedings of the ACM Web Conference 2024}{May 13--17, 2024}{Singapore, Singapore} \acmBooktitle{Companion Proceedings of the ACM Web Conference 2024 (WWW '24 Companion), May 13--17, 2024, Singapore, Singapore}\acmDOI{10.1145/3589335.3651555}
\acmISBN{979-8-4007-0172-6/24/05}

\makeatletter
\gdef\@copyrightpermission{
  \begin{minipage}{0.3\columnwidth}
   \href{https://creativecommons.org/licenses/by/4.0/}{\includegraphics[width=0.90\textwidth]{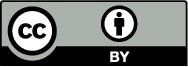}}
  \end{minipage}\hfill
  \begin{minipage}{0.7\columnwidth}
   \href{https://creativecommons.org/licenses/by/4.0/}{This work is licensed under a Creative Commons Attribution International 4.0 License.}
  \end{minipage}
  \vspace{5pt}
}
\makeatother

\begin{document}

\title{Robust Federated Learning Mitigates Client-side Training Data Distribution Inference Attacks}

\author{Yichang Xu}
\authornote{Equal contribution.}
\affiliation{%
  \institution{University of Science and Technology of China}
}
\email{xuyichang@mail.ustc.edu.cn}

\author{Ming Yin}
\authornotemark[1]
\affiliation{%
	\institution{University of Science and Technology of China}
}
\email{mingyin@mail.ustc.edu.cn}

\author{Minghong Fang}
\affiliation{%
	\institution{Duke University}
}
\email{minghong.fang@duke.edu}

\author{Neil Zhenqiang Gong}
\affiliation{%
	\institution{Duke University}
}
\email{neil.gong@duke.edu}

\begin{abstract}	 

\input{abstract}

\end{abstract}



\maketitle

\input{introduction}

\input{threatModel}

\input{method}

\input{experiments}

\input{conclusion}

\balance
\bibliographystyle{ACM-Reference-Format}
\bibliography{refs}

\end{document}

%% file: abstract.tex
Recent studies have revealed that federated learning (FL), once considered secure due to clients not sharing their private data with the server, is vulnerable to attacks such as client-side training data distribution inference, where a malicious client can recreate the victim's data. While various countermeasures exist, they are not practical, often assuming server access to some training data or knowledge of label distribution before the attack.

In this work, we bridge the gap by proposing InferGuard, a novel Byzantine-robust aggregation rule aimed at defending against client-side training data distribution inference attacks. In our proposed InferGuard, the server first calculates the coordinate-wise median of all the model updates it receives. A client's model update is considered malicious if it significantly deviates from the computed median update. We conduct a thorough evaluation of our proposed InferGuard on five benchmark datasets and perform a comparison with ten baseline methods. The results of our experiments indicate that our defense mechanism is highly effective in protecting against client-side training data distribution inference attacks, even against strong adaptive attacks. Furthermore, our method substantially outperforms the baseline methods in various practical FL scenarios.

%% file: introduction.tex

\section{Introduction} \label{sec:intro}
Federated learning (FL)~\cite{McMahan2016CommunicationEfficientLO} is an innovative distributed machine learning paradigm that has gained significant attention recently, allowing individuals to train a global machine learning model collaboratively without sharing their private training data with others.
FL usually consists of one server and multiple clients. During the FL training process, each client performs local training using the current global model and its local training data, then sends its local model update to the server. 
Upon receiving model updates from all clients, the server leverages a specific aggregation rule to combine the received model updates and further update the global model.
The updated global model is then distributed to all clients for the next round of training. 
For instance, the FedAvg~\cite{McMahan2016CommunicationEfficientLO} aggregation rule calculates the average of model updates to obtain the global model and is commonly employed in non-adversarial scenarios. 

A benefit of FL compared with centralized learning is that clients no longer need to send their private training data to the server. Some cryptographic approaches have been proposed to protect the FL system from information leakage in network security level~\cite{bonawitz2017practical}.
However, methods reliant on cryptography often incur large computation and communication overhead~\cite{zhang2020batchcrypt}.
Recent studies~\cite{fang2020local,baruch2019little} 
have shown that FL is vulnerable to poisoning attacks, where malicious clients could send carefully crafted model updates to the server to manipulate the final learned global model. Moreover, some works studied the privacy of FL and found that FL is not as sound as expected in privacy--even awful. These works explored the possibility of privacy both on the server and client sides. On the server side, they found that the server can reconstruct images or properties of data from a specific client~\cite{yue2023gradient}.
The most influential for client-side inference attacks is the GAN attack proposed by Hitaj et al.~\cite{hitaj2017deep}. 
Taking advantage of the dynamic nature of the learning process, this form of attack enables adversaries to leverage a Generative Adversarial Network (GAN)~\cite{goodfellow2014generative} to produce similar samples from the targeted training set, initially intended to remain undisclosed. The models generated by the GAN aim to align closely with the same data distribution as the training data.

Client-side inference attacks are considerably more feasible than server-side ones, as the attacker involved in a client-side attack is a participating client in FL, and does not need to compromise the server to carry out its malicious actions.
Certain defense mechanisms have been specifically designed to defend against client-side inference attacks. For instance, Netzer et al. \cite{wei2021gradientleakage} developed a method that utilizes differential privacy mechanisms to mitigate such attacks. 
Nevertheless, our later experiments revealed that such approaches are ineffective in defending against inference attacks.

\myparatight{Our work} We first observe that existing Byzantine-robust aggregation rules~\cite{blanchard2017machine,mhamdi2018hidden,yin2021byzantinerobust,fang2022aflguard,cao2022fltrust} could only mitigate client-side inference attacks to some extent. However, they are still vulnerable to inference attacks in certain circumstances, as demonstrated in our experiments. Based on this motivation, we propose a new Byzantine-robust aggregation rule called \alg{} to defend against inference attacks. 
In our proposed \alg{}, after receiving model updates from all clients, the server calculates the Median~\cite{yin2021byzantinerobust} of these updates. If a received model update deviates substantially from the computed Median, it is identified as malicious.

\myparatight{\alg{} can defend against client-side training data distribution inference attacks} We extensively evaluate our proposed \alg{} on five datasets, including MNIST~\cite{lecun2010mnist}, Fashion-MNIST~\cite{xiao2017/online}, AT\&T~\cite{Samaria1994ParameterisationOA}, SVHN~\cite{Netzer2011ReadingDI}, and GTSRB~\cite{Stallkamp-IJCNN-2011}.
We compared \alg{} with several existing Byzantine-robust aggregation rules, including AFA~\cite{munoz2019byzantine}, Multi-Krum~\cite{blanchard2017machine}, Bulyan~\cite{mhamdi2018hidden}, Trimmed mean~\cite{yin2021byzantinerobust}, Median~\cite{yin2021byzantinerobust}, FLTrust~\cite{cao2022fltrust}, along with model update post-processing mechanisms such as compression~\cite{bernstein2018signsgd}, sparsification~\cite{aji2017sparse}, and differential privacy~\cite{wei2021gradientleakage}. Our findings demonstrate the robustness of \alg{} against client-side inference attacks, whereas the aforementioned Byzantine-robust aggregation rules and other defense mechanisms exhibit vulnerability in specific scenarios.

Our contributions can be summarized as follows:
\begin{itemize}
 
 \item We find that existing Byzantine-robust aggregation rules could mitigate the client-side inference attacks on FL to some extent. However, their effect is not optimal because we can still recognize the content in generated images.

 \item We introduce \alg{}, an innovative defense framework designed to protect against inference attacks on the client side of FL. \alg{} effectively mitigates malicious clients' influence while preserving the FL system's utility.

\item We thoroughly evaluate our defense framework using five benchmark datasets. Our results demonstrate that \alg{} effectively safeguards against client-side inference attacks, and outperforms baseline methods.

\end{itemize}

%% file: threatModel.tex

\section{Threat Model and Defense Goals} \label{sec:threat_model}

\myparatight{Attacker's Goal} 
Following~\cite{hitaj2017deep}, we consider a client-side data reconstruction attack. We assume that an attacker controls some malicious clients.
The attacker's goal is to reconstruct images of a specific label that it initially did not possess.

\myparatight{Attacker's Capabilities} 
The attacker achieves his goal via sending carefully crafted model updates to the server.
Malicious clients can exchange local model updates amongst themselves, and each malicious client knows other malicious clients' local training data. Malicious clients can initiate the attack at any global round. 
Note that the server is trustworthy, and there is no collusion between the attacker and the server.

\myparatight{Defense Goals} 
Our goal is to create an aggregation rule for FL systems that is secure against client-side inference attacks and maintains robustness, fidelity, and efficiency. This rule should prevent attackers from accessing clients' local data, ensure the global model's accuracy is as close to FedAvg's as possible without attacks, and be computationally efficient without additional costs.

%% file: method.tex

\section{Our Method}

Suppose we have $n$ clients, in global training round $t$, each client $i$ submits its local model update $\bm{g}_i^t$ to the server, where $1 \le i \le n$.
Upon receiving local model updates from all clients, the server first computes the coordinate-wise median  of $n$ local model updates, denoted as $\bm{g}_{\text{med}}^t$, as the following:
\begin{align}
    \bm{g}_{\text{med}}^t = \text{Median} \{\bm{g}_1^t, \bm{g}_2^t,..., \bm{g}_n^t \},
\end{align}
where $\text{Median} \{\bm{g}_1^t, \bm{g}_2^t,..., \bm{g}_n^t \}$ is the coordinate-wise median~\cite{yin2021byzantinerobust} of these $n$ model updates.

Subsequently, the server determines whether $\bm{g}_i^t$ should be considered a benign local model update based on the following:
\begin{align}
\label{our_agg}
\left\| \bm{g}_i^t -  \bm{g}_{\text{med}}^t \right\|_2 \le \lambda  \left\| \bm{g}_{\text{med}}^t   \right\|_2.
\end{align}

During the global training round $t$,
the set of clients whose model updates satisfy Eq.~(\ref{our_agg}) is denoted as $\mathcal{H}$. The final aggregated local model update is then calculated as the average of the local model updates of all the clients in $\mathcal{H}$ as $\frac{1}{\left|  \mathcal{H} \right|} \sum\nolimits_{i \in \mathcal{H}} \bm{g}_i^t $. If $|\mathcal{H}|=0$, then we choose the model update whose $\left\| \bm{g}_i^t -  \bm{g}_{\text{med}}^t \right\|_2$ is the smallest.

\begin{figure}[!t]
	\centering
	\subfloat[Multi-Krum]
	{\includegraphics[width=0.2 \textwidth]{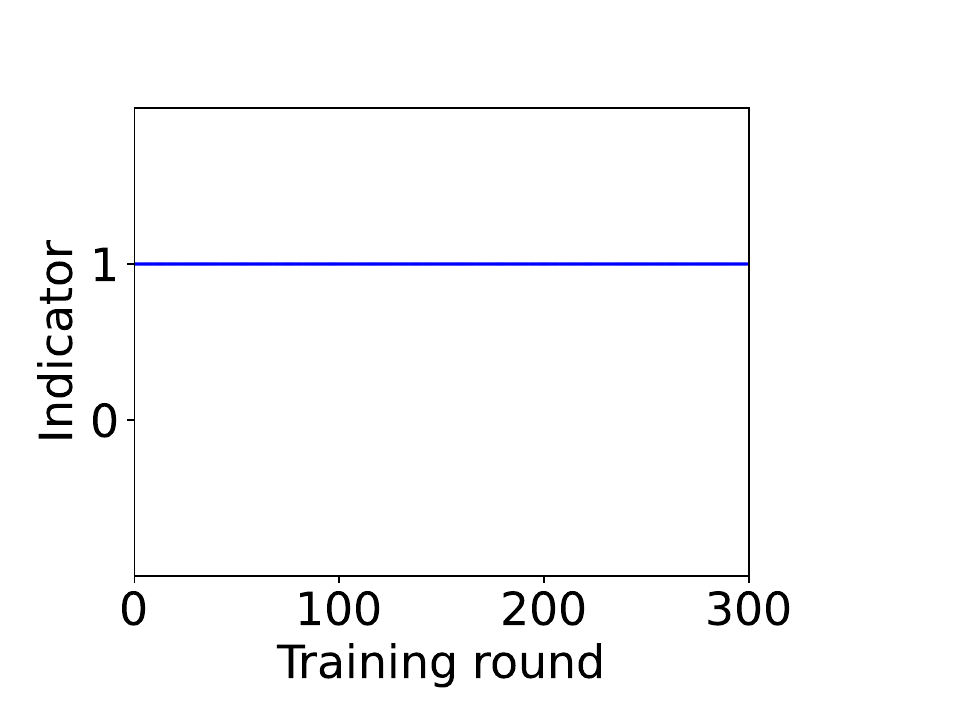}}
    \subfloat[\alg{}]
	{\includegraphics[width=0.2 \textwidth]{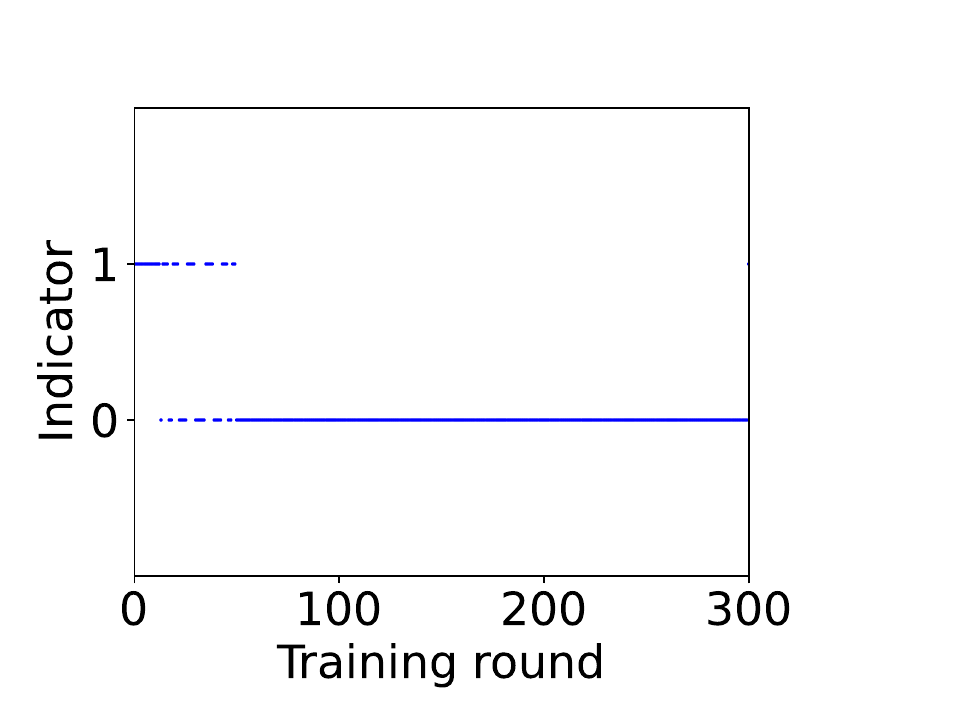}}
	\caption{MNIST dataset, whether a malicious model update is chosen in each round.}
		\label{mnist_indicator}
    \vspace{-3mm}
\end{figure}

We experiment on the MNIST dataset to verify our idea (refer to Section~\ref{Experimental_Setup} for experimental settings).
We aim to check whether the model update from the malicious client is chosen at each global round. Multi-Krum is adopted as our baseline. 
The results are shown in Figure \ref{mnist_indicator}.
In Figure \ref{mnist_indicator}, if the ``Indicator'' equals one, it represents that the server mistakenly selects the malicious model update in a specific training round. In contrast, zero represents that it does not. 
From Figure \ref{mnist_indicator}, we observe that for Multi-Krum, the server chooses the malicious model update almost every training round. However, in our method, after the attacker starts to attack at the $50$th training round (following prior work~\cite{hitaj2017deep}, we assume that the attacker starts to attack from a specific training round), the server does not select the malicious model update anymore.

%% file: experiments.tex

\section{Experiments} \label{sec:exp}

\subsection{Experimental Setup}
\label{Experimental_Setup}
\subsubsection{Datasets} 
We use MNIST~\cite{lecun2010mnist}, Fashion-MNIST~\cite{xiao2017/online}, AT\&T~\cite{Samaria1994ParameterisationOA}, SVHN~\cite{Netzer2011ReadingDI} and GTSRB~\cite{Stallkamp-IJCNN-2011} datasets in our experiments.

\subsubsection{Evaluation Metrics}Following~\cite{geiping2020inverting,zhu2019deep,yue2023gradient}, we use Learned Perceptual Image Patch Similarity (LPIPS)~\cite{zhang2018unreasonable}, Peak Signal-to-Noise Ratio (PSNR)~\cite{castleman1996digital}, and Structural Similarity Index Measure (SSIM)~\cite{wang2004image} as evaluation metrics. Larger LPIPS or lower PSNR/SSIM indicates better defense performance. According to~\cite{nilsson2020understanding}, SSIM can be considered as the primary index to measure the defense effect.

\subsubsection{Compared Methods} 
We compare with our proposed \alg{} with 10 baselines, includes 7 aggregation rules (FedAvg~\cite{McMahan2016CommunicationEfficientLO}, adaptive federated average (AFA)~\cite{munoz2019byzantine}, Multi-Krum~\cite{blanchard2017machine}, Bulyan~\cite{mhamdi2018hidden}, coordinate-wise trimmed mean (Trim)~\cite{yin2021byzantinerobust}, coordinate-wise median (Median)~\cite{yin2021byzantinerobust}, FLTrust~\cite{cao2022fltrust}), and 3 post-processing defenses (sparsification~\cite{aji2017sparse}, compression~\cite{bernstein2018signsgd}, differential privacy (DP)~\cite{wei2021gradientleakage}).

\subsubsection{Parameter Settings} 
We train 300 rounds on MNIST, Fashion-MNIST, AT\&T, GTSRB datasets, and 150 rounds on SVHN, with each client training locally for one epoch per round. In a FL setup with 10 clients, where one is malicious following the approach in \cite{hitaj2017deep}, we simulate non-i.i.d. data by uniformly distributing each label to 5 clients, leading to distinct class distributions per client. In all datasets, we suppose that the last client is the malicious client. In MNIST, Fashion-MNIST, GTSRB, and SVHN datasets, the malicious client steals images of label 3, which he does not own. In AT\&T, the malicious client steals images of label 11. The malicious client initiates data distribution inference attacks targeting specific labels not owned by them, starting from round 50 for MNIST, Fashion-MNIST, AT\&T, and GTSRB, and round 20 for SVHN. For the $\lambda$ parameter, we set $\lambda=2.0$ for MNIST, $\lambda=1.8$ for Fashion-MNIST, $\lambda=2.8$ for AT\&T, $\lambda=3.0$ for SVHN and $\lambda=1.2$ for GTSRB dataset.

\begin{table*}[htbp]
 \scriptsize
  \centering
  \caption{Results of different FL methods under attack.}
  \subfloat[MNIST]{
    \begin{tabular}{|c|c|c|c|c|c|c|c|c|c|c|c|c|}
    \hline
          & No attack & FedAvg  & AFA  & Multi-Krum & Bulyan & Trim  & Median & FLTrust & Sparsification & Compression & DP    & \alg{} \\
    \hline
    LPIPS  $\uparrow$  & 0.00 & 0.34 &  0.32 & 0.28 & 0.27 & 0.36 & 0.36 & 0.35 & 0.34 & 0.35 & 0.31 & 0.39 \\
    \hline
    PSNR  $\downarrow$ & inf & 28.55 & 28.70 & 29.97 & 31.05 & 28.91 & 29.28 & 28.76 & 28.51 & 28.60 & 28.76 & 28.53 \\
    \hline
    SSIM  $\downarrow$ & 1.00 & 0.53 & 0.55 & 0.66 & 0.59 & 0.53 & 0.57 & 0.40 & 0.56 & 0.55 & 0.57 & 0.22 \\
    \hline
    \end{tabular}%
    }

  \subfloat[Fashion-MNIST]{
    \begin{tabular}{|c|c|c|c|c|c|c|c|c|c|c|c|c|}
    \hline
          & No attack & FedAvg  & AFA  & Multi-Krum & Bulyan & Trim  & Median & FLTrust & Sparsification & Compression & DP    & \alg{} \\
    \hline
    LPIPS  $\uparrow$  & 0.00 & 0.34 & 0.34 & 0.35 & 0.34 & 0.32 & 0.35 & 0.34 & 0.34 & 0.35 & 0.33 & 0.44 \\
    \hline
    PSNR  $\downarrow$ & inf & 29.75 & 30.16 & 29.46 & 29.00 & 29.58 & 29.56 & 29.53 & 29.8 & 29.44 & 29.77 & 28.94 \\
    \hline
    SSIM  $\downarrow$ & 1.00 & 0.34 & 0.35 & 0.28 & 0.33 & 0.37 & 0.35 & 0.3 & 0.36 & 0.31 & 0.35 & 0.21 \\
    \hline
    \end{tabular}%
    }

  \subfloat[AT\&T]{
    \begin{tabular}{|c|c|c|c|c|c|c|c|c|c|c|c|c|}
    \hline
          & No attack & FedAvg  & AFA  & Multi-Krum & Bulyan & Trim  & Median & FLTrust & Sparsification & Compression & DP    & \alg{} \\
    \hline
    LPIPS  $\uparrow$  & 0.00 & 0.30 & 0.32 & 0.30 & 0.30 & 0.31 & 0.31 & 0.31 & 0.29 & 0.29 & 0.31 & 0.28 \\
    \hline
    PSNR  $\downarrow$ & inf & 28.10 & 28.10 & 28.27 & 28.19 & 28.13 & 28.16 & 28.15 & 28.18 & 28.29 & 28.13 & 28.15 \\
    \hline
    SSIM  $\downarrow$ & 1.00 & 0.31 & 0.30 & 0.41 & 0.44 & 0.30 & 0.34 & 0.27 & 0.35 & 0.36 & 0.34 & 0.21 \\
    \hline
    \end{tabular}%
    }

    \subfloat[SVHN]{
    \begin{tabular}{|c|c|c|c|c|c|c|c|c|c|c|c|c|}
    \hline
          & No attack & FedAvg  & AFA  & Multi-Krum & Bulyan & Trim  & Median & FLTrust & Sparsification & Compression & DP    & \alg{} \\
    \hline
    LPIPS  $\uparrow$  & 0.00 & 0.14 & 0.13 & 0.11 & 0.11 & 0.12 & 0.20 & 0.13 & 0.13 & 0.27 & 0.20 & 0.18 \\
    \hline
    PSNR  $\downarrow$ & inf & 28.68 & 28.65 & 28.81 & 28.97 & 28.66 & 28.67 & 28.73 & 28.95 & 28.76 & 28.95 & 28.68 \\
    \hline
    SSIM  $\downarrow$ & 1.00 & 0.62 & 0.58 & 0.65 & 0.64 & 0.66 & 0.57 & 0.65 & 0.63 & 0.41 & 0.59 & 0.53 \\
    \hline
    \end{tabular}%
    }

    \subfloat[GTSRB]{
    \begin{tabular}{|c|c|c|c|c|c|c|c|c|c|c|c|c|}
    \hline
          & No attack & FedAvg  & AFA  & Multi-Krum & Bulyan & Trim  & Median & FLTrust & Sparsification & Compression & DP    & \alg{} \\
    \hline
    LPIPS  $\uparrow$  & 0.00 & 0.25 & 0.29 & 0.24 & 0.25 & 0.24 & 0.22 & 0.27 & 0.25 & 0.28 & 0.22 & 0.32 \\
    \hline
    PSNR  $\downarrow$ & inf & 29.99 & 29.99 & 29.83 & 28.29 & 29.75 & 29.38 & 28.51 & 30.00 & 29.97 & 30.00 & 28.59 \\
    \hline
    SSIM  $\downarrow$ & 1.00 & 0.19 & 0.21 & 0.13 & 0.12 & 0.19 & 0.13 & 0.12 & 0.25 & 0.19 & 0.20 & 0.02 \\
    \hline
    \end{tabular}%
    }
  \label{tab:default_eval}%
\end{table*}%

\begin{figure}[!t]
	\centering
     \subfloat[No attack]
	{\includegraphics[width=0.16 \textwidth]{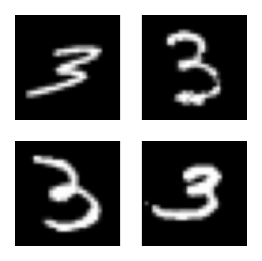}}
    \subfloat[Multi-Krum]
	{\includegraphics[width=0.16 \textwidth]{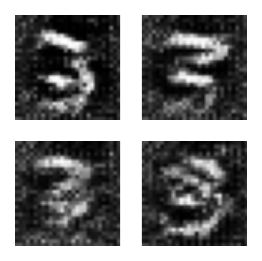}}
	\subfloat[\alg{}]
	{\includegraphics[width=0.16 \textwidth]{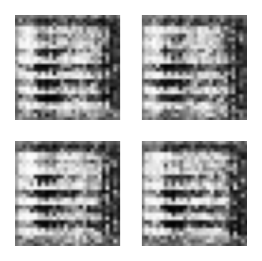}}
	\caption{Reconstruction samples on MNIST dataset.}
		\label{mnist_5label}
  	\vspace{-3mm}
\end{figure}

\subsection{Experimental Results}
\label{Experimental_Results}

\myparatight{Our \alg{} is effective} 
From Table~\ref{tab:default_eval}, we observe that our method almost wins on all three evaluation metrics compared with baselines on five datasets. For instance, the SSIM value on MNIST after our defense declined to 0.22, while SSIM values of baselines are at least 0.4. Our visualization result, Figure~\ref{mnist_5label} also supports the same conclusion. 

\begin{figure}[!t]
	\centering
	\subfloat[LPIPS]
	{\includegraphics[width=0.16 \textwidth]{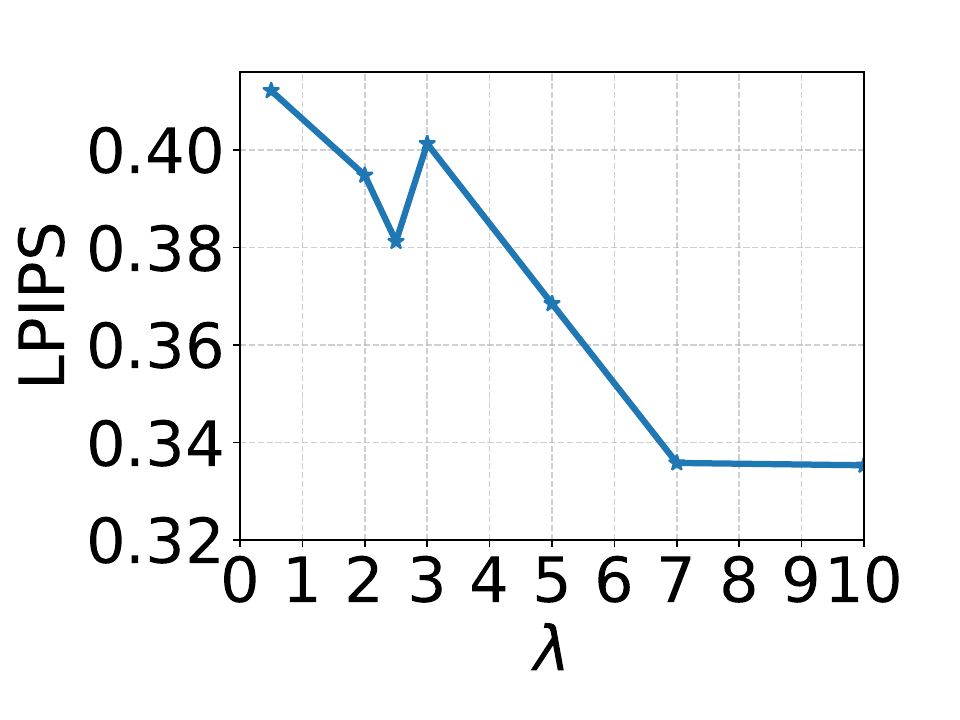}}
	\subfloat[PSNR]
	{\includegraphics[width=0.16 \textwidth]{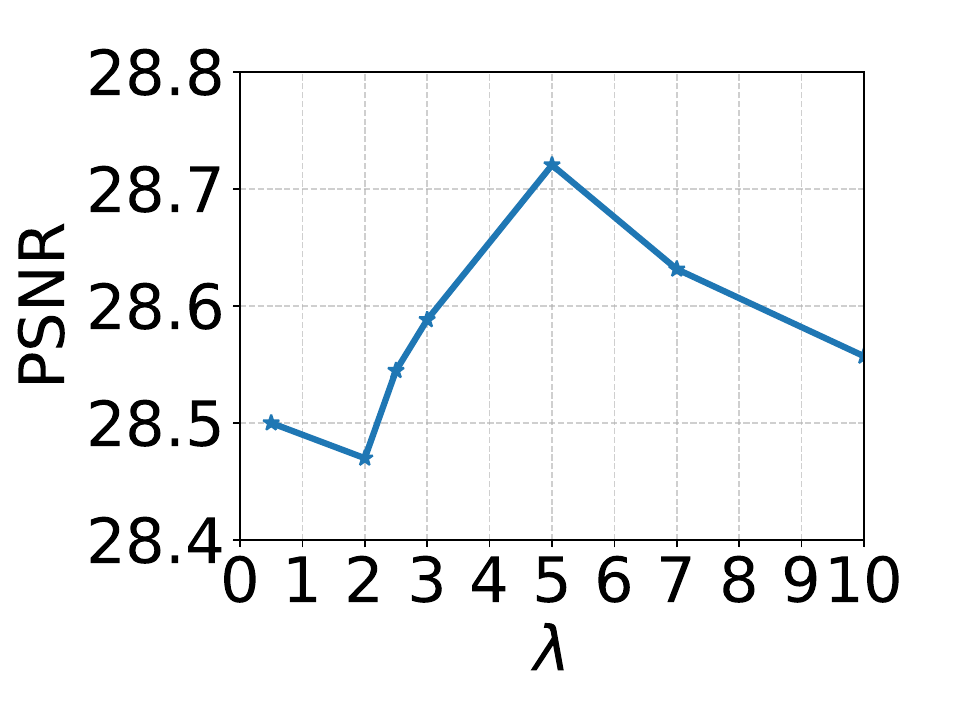}}
	\subfloat[SSIM]
	{\includegraphics[width=0.16 \textwidth]{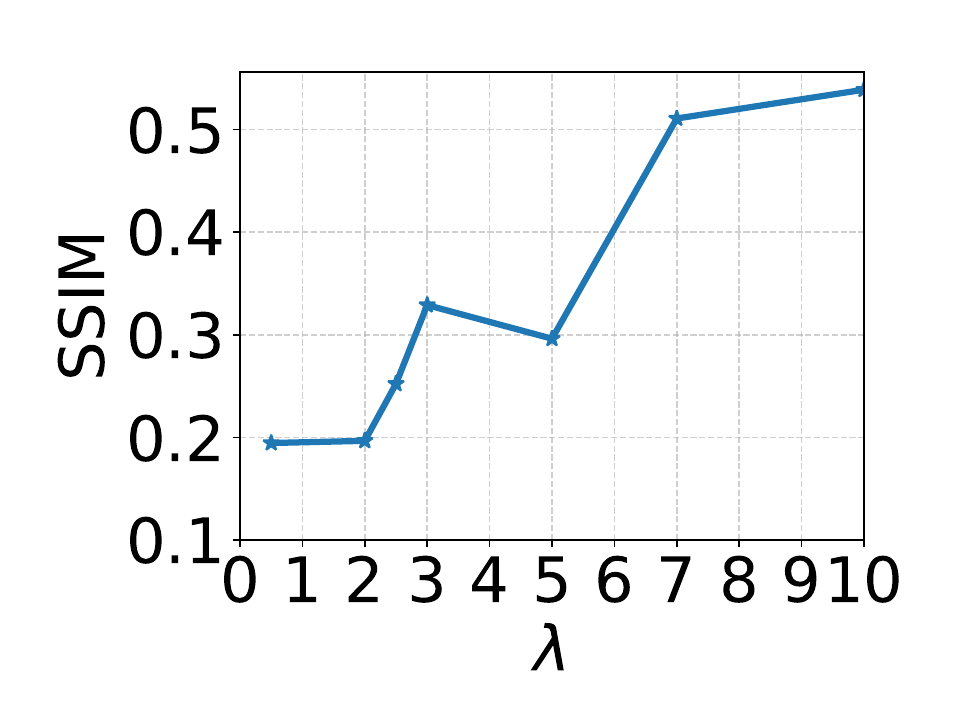}}
	\caption{Impact of different $\lambda$ on MNIST dataset.}
		\label{diff_lambda_eval}
   \vspace{-3mm}
\end{figure}

\myparatight{Impact of different $\lambda$}
We experiment with the parameter $\lambda$ taking on values of $0.5, 2, 2.5, 3, 5, 7, 10$. As shown in Figure~\ref{diff_lambda_eval}, we can conclude that the defensive efficacy starts to decline when $\lambda$ surpasses 2. So in our experiment, we choose $\lambda=2$ as the default setting. 

\myparatight{Non-iid settings}
Table~\ref{tab:diff_label} shows the situations where each client owns three labels. Under such non-iid settings, our method is still superior over baselines.

\myparatight{Results on adaptive attacks}
Suppose the local model weight is $\bm{w}_i^t$, and the loss function when training the local model is $L(\bm{w}_i^t;\bm{x})$. The training data distribution for the malicious client is $\mathcal{D}_i$. The optimization problem for the malicious client is formalized as $\min_{\bm{w}_i^t}\mathbb{E}_{\bm{x}\sim\mathcal{D}_i}[L(\bm{w}_i^t;\bm{x})], \quad s.t. \|\bm{w}_i^t-\bm{w}^t\|_{\infty}\le \tau$, where $\tau$ is a constant. Table \ref{tab:adaptive_eval} shows the evaluation result of our adaptive attack on the MNIST dataset with $\tau=0.0016$. In MNIST, our defense effect is weakened, but from the SSIM values, we can conclude that it is still robust in some way because the SSIM score under our method is still lower than SSIM scores under baselines. 

\myparatight{Results on membership inference attack}
We consider the membership inference attack proposed in~\cite{zhang2023agrevader}. The attacker has 300 samples and needs to determine whether each sample is in the training set of the other 9 clients. Other settings are the same as our default setting. Table~\ref{tab:meminfer} displays the results on Location30~\cite{shokri2017membership} dataset, as recommended in~\cite{zhang2023agrevader}. In Table~\ref{tab:meminfer}, ``Model acc'' denotes the global model's testing accuracy, and ``Attack acc'' represents the attacker's success rate. We can see that \alg{} achieves the best defense effect among all defenses. Moreover, it continues to sustain the model's high-level performance.

\begin{table*}[htbp]
  \scriptsize
  \centering
  \caption{Results of adaptive attack on MNIST dataset.}
    \begin{tabular}{|c|c|c|c|c|c|c|c|c|c|c|c|c|}
    \hline
          & No attack & FedAvg  & AFA  & Multi-Krum & Bulyan & Trim  & Median & FLTrust & Sparsification & Compression & DP    & \alg{} \\
    \hline
    LPIPS  $\uparrow$  & 0.00 & 0.37 & 0.34 & 0.31 & 0.28 & 0.31 & 0.32 & 0.33 & 0.32 & 0.33 & 0.31 & 0.38 \\
    \hline
    PSNR  $\downarrow$ & inf & 28.52 & 28.61 & 29.34 & 30.26 & 29.15 & 28.95 & 29.05 & 28.73 & 28.81 & 28.72 & 28.68 \\
    \hline
    SSIM  $\downarrow$ & 1.00 & 0.50 & 0.54 & 0.59 & 0.59 & 0.62 & 0.57 & 0.54 & 0.54 & 0.56 & 0.55 & 0.48 \\
    \hline
    \end{tabular}
  \label{tab:adaptive_eval}%
\end{table*}%

\begin{table*}[htbp]
   \scriptsize
  \centering
  \caption{Impact of data distribution. Each client owns three labels of data.}
    
\begin{tabular}{|c|c|c|c|c|c|c|c|c|c|c|c|c|}
    \hline
          & No attack & FedAvg  & AFA  & Multi-Krum & Bulyan & Trim  & Median & FLTrust & Sparsification & Compression & DP    & \alg{} \\
    \hline
    LPIPS  $\uparrow$  & 0.00 & 0.32 & 0.36 & 0.27 & 0.28 & 0.33 & 0.45 & 0.30 & 0.36 & 0.34 & 0.34 & 0.42 \\
    \hline
    PSNR  $\downarrow$ & inf & 28.9 & 28.95 & 29.68 & 29.59 & 29.11 & 28.84 & 29.75 & 28.56 & 29.71 & 28.90 & 28.54 \\
    \hline
    SSIM  $\downarrow$ & 1.00 & 0.50 & 0.54 & 0.65 & 0.62 & 0.57 & 0.33 & 0.60 & 0.52 & 0.50 & 0.55 & 0.27 \\
    \hline
    \end{tabular}%
  \label{tab:diff_label}%
\end{table*}%

\begin{table*}[htbp]
 \scriptsize
  \centering
  \caption{Results of different FL methods under membership inference attack on Location30 dataset.}
    \begin{tabular}{|c|c|c|c|c|c|c|c|c|c|c|c|c|}
    \hline
          & FedAvg  & AFA & Multi-Krum & Bulyan & Trim  & Median & FLTrust & Sparsification & Compression & DP    & \alg{} \\
    \hline
     Model acc  $\uparrow$  & 0.65  & 0.64 & 0.58 & 0.65 & 0.69 & 0.68 & 0.72 & 0.68 & 0.69 & 0.65 & 0.70 \\     \hline
    Attack acc $\downarrow$  & 0.70  & 0.67 & 0.66 & 0.63 & 0.66 & 0.65 & 0.63 & 0.65 & 0.65 & 0.69 & 0.61 \\

    \hline
    \end{tabular}%
  \label{tab:meminfer}%
     \vspace{-1.5mm}
\end{table*}%

%% file: conclusion.tex

\section{Conclusion} \label{sec:conclusion}
Our research demonstrated that current strategies for defending against client-side inference attacks fall short in practice, highlighting the need for a more robust defense mechanism in FL. 
Interestingly, we discovered that existing Byzantine-robust aggregation rules, although not originally designed to combat inference attacks, provide a degree of protective effect.
Building on these insights, we developed a novel Byzantine-robust aggregation rule named \alg{}, which could effectively counter client-side inference attacks. Comparative analysis across five datasets revealed that this new defense strategy markedly outperforms existing mechanisms. Future exploration could focus on providing a formal theoretical guarantee to demonstrate the robustness of our proposed \alg{} against client-side training data distribution inference attacks.

\begin{acks}
We thank the anonymous reviewers for their comments. 
This work was supported by NSF grant No. 2131859, 2125977, 2112562, 1937786, 1937787, and ARO grant No. W911NF2110182.
\end{acks}